\documentclass[conference]{IEEEtran}
\usepackage[english]{babel}
\usepackage[utf8]{inputenc}
\usepackage[T1]{fontenc}

\usepackage[compact]{titlesec}
\titlespacing{\section}{.6pt}{*.6}{*.6}
\titlespacing{\subsection}{0.6pt}{*0.6}{*0.6}
\titlespacing{\subsubsection}{0.5pt}{*0.5}{*0.5}

\usepackage{graphicx}
\usepackage{subfigure}
\usepackage{float}
\usepackage{wrapfig}
\usepackage{titlesec}

\usepackage{caption}
\usepackage{amsmath}
\usepackage{nth}
\usepackage{soul}
\usepackage{listings}
\usepackage[autostyle]{csquotes}
\usepackage[shortlabels, inline]{enumitem}

\usepackage[hidelinks]{hyperref}
\hypersetup{colorlinks=true,linkcolor=blue,citecolor=blue,urlcolor=blue}

\usepackage{color, xcolor} 
\usepackage{array, tabularx, makecell, multirow, booktabs}
\newcolumntype{L}[1]{>{\raggedright\arraybackslash}p{#1}}
\newcolumntype{Z}{>{\centering\let\newline\\\arraybackslash\hspace{0pt}}X}
\newcolumntype{Y}{>{\raggedright\let\newline\\\arraybackslash\hspace{0pt}}X}

\newcommand{\subhead}[1]{\noindent{\textbf{#1:}}} 

\usepackage{xspace}
\newcommand{\eg}{\textit{e.g.,}\xspace}
\newcommand{\ie}{\textit{i.e.,}\xspace}

\newcommand{\sol}{\textsc{FlowGuardian}\xspace}

\newcommand{\faissal}{\color{black}}
\newcommand{\sepehr}{\color{black}}
\newcommand{\ahmadou}{\color{black}}
\newcommand{\revision}{\color{black}}

\begin{document}
\bstctlcite{IEEEexample:BSTcontrol}

\title{Automated Attack Testflow Extraction from Cyber Threat Report using BERT for Contextual Analysis}

\author{
    \IEEEauthorblockN{
        Faissal Ahmadou\IEEEauthorrefmark{1},
    	Sepehr Ghaffarzadegan\IEEEauthorrefmark{1},
        Boubakr Nour\IEEEauthorrefmark{4}, 
        Makan Pourzandi\IEEEauthorrefmark{4},
        Mourad Debbabi\IEEEauthorrefmark{1}, 
        Chadi Assi\IEEEauthorrefmark{1}
    }
    \IEEEauthorblockA{\IEEEauthorrefmark{1}Concordia University, Canada
    \qquad
    \IEEEauthorrefmark{4}Ericsson Security Research, Canada}
}

\maketitle

\begin{abstract}
    In the ever-evolving landscape of cybersecurity, the rapid identification and mitigation of Advanced Persistent Threats (APTs) is crucial. Security practitioners rely on detailed threat reports to understand the tactics, techniques, and procedures (TTPs) employed by attackers. However, manually extracting attack testflows from these reports requires elusive knowledge and is time-consuming and prone to errors.
    This paper {\faissal proposes} \sol, a novel solution leveraging language models (\ie BERT) and Natural Language Processing (NLP) techniques to automate the extraction of attack testflows from unstructured threat reports. \sol systematically analyzes and contextualizes security events, reconstructs attack sequences, and {\faissal then} generates {\faissal comprehensive} testflows. This automated approach not only saves time and reduces human error but also ensures comprehensive coverage and robustness in cybersecurity testing.
    Empirical validation using public threat reports demonstrates \sol's accuracy and efficiency, significantly enhancing the capabilities of security teams in proactive threat hunting and incident response.
\end{abstract}


\IEEEpeerreviewmaketitle

\section{Introduction}
\label{sec:introduction}
%
{\sepehr Advanced Persistent Threats (APTs)~\cite{alshamrani2019survey} pose a complex security challenge due to their stealthy nature.
} 
{\sepehr APTs execute multi-stage campaigns, using \textit{Living off the Land} strategies to evade detection and persist.}
To counteract these threats, threat intelligence companies (\eg Mandiant, CrowdStrike) and organizations meticulously document their encounters with APTs in threat reports\footnote{Throughout this paper, the term \textit{threat report} is used broadly to describe any document that details a threat (\eg cyber threat reports, security/assessment reports, security troubleshooting reports).}. These reports detail the specific observed campaigns, including the tactics, techniques, and procedures (TTPs) employed by the attackers. 
%
For example, Mandiant has extensively documented APT41's activities, detailing a specific campaign exploiting CVE-2019-3396 in~\cite{googleblog2024}, a campaign leveraging CVE-2019-19781 in~\cite{googleblog2023}, and the attack TTPs alongside the vulnerabilities the threat actor exploited in~\cite{mandiant2024}.
%
%
%
Moreover, Google has offered insights into new APT41 variants~\cite{gcat2023apr} which utilize Google C2C, Google Sheets, and Google Drive.
Each of these reports provides crucial observables from the respective campaigns.

Understanding those reports is crucial to security practitioners, especially for extracting testflow (\eg security playbook) to verify if the organization is under such an attack. A testflow is a structured set of information that helps practitioners understand the sequence of steps an attacker takes during a cyber attack.
Security practitioners start by thoroughly reading and understanding the report to grasp the scope, nature of the threat, affected systems, and exploitation details. 
After identifying the attack steps and objectives, 
they extract critical technical details (\eg specific software versions, configurations, and necessary conditions for the threat) which will guide the development of a detailed testflow.
The process is inherently complex and demands not only advanced technical knowledge and analytical skills which are elusive, but also a significant time investment. The process becomes more complicated and cumbersome when multiple threat reports are released leading to further efforts to generate a comprehensive testflow.
Therefore, automating testflow extraction from threat reports helps in ensuring consistency, accuracy, and efficiency. By implementing automation, security practitioners can parse through extensive and diverse reports coming from different sources simultaneously, identify common vulnerabilities that might lead to the attack, and consolidate similar testflow. This not only saves time and reduces the effort required by security teams but also minimizes human error and enhances the thoroughness of the testing process. 
%

{\sepehr Generated testflows serve as essential resources for {\revision Security Operation Center} (SOC) teams:}
\begin{enumerate*}[(i)]
    \item they enable proactive threat hunting by providing a framework to systematically search for signs of APTs (\eg Indicators of Compromise -- IoCs) that match the behaviors described in the threat reports; and 
    \item the testflows facilitate a more informed and efficient incident response. 
    When an attack is detected, SOC teams can use the specific testflows relevant to the observed attack to quickly understand the nature of the threat and trace its origin and impact.
\end{enumerate*}

Unlike existing studies that concentrate on deriving attack techniques from threat reports (\eg~\cite{li2024automated}) or from CVEs (\eg~\cite{abdeen2023smet}), our work primarily targets the extraction of attack testflows from previously observed attack campaigns documented in threat reports. This approach aims to enrich the cybersecurity playbook. Thus, in this paper, we make the following contributions:
{\revision
\begin{enumerate*}[(1)]
    \item we introduce a novel solution, \sol, tailored specifically for extracting testflows from threat reports (unstructured documents). Our solution is the first that uses language models to perform security contextual analysis and sequence analysis to systematically break down the attack into actionable steps and thus generate an accurate attack testflow.
    \item we design an automated framework that integrates entity extraction and, sequence analysis, enabling efficient extraction of testflows from raw text without manual intervention.
    \item we validate our solution through empirical studies using public APT threat reports. The obtained results, reviewed and confirmed
    by security practitioners, show that \sol is able to extract accurate testflows and enrich the cybersecurity playbook.
\end{enumerate*}
} 
%


\section{Motivation and Problem Statement}
\label{sec:background}
\subhead{Motivation}
Whenever a new threat report is released, 
security analysts have a critical responsibility to review the document and assess whether their organization is subject to the reported attack. This process underscores the importance of testflow extraction from security reports. To effectively, yet proactively, secure their networks/systems, they must thoughtfully identify the attack infrastructure outlined in the report and meticulously construct the attack sequence and steps. This enables SOC teams to test if their organization is vulnerable to the documented attack steps.  The ability to quickly and accurately extract attack testflow is crucial for proactive defense measures and ensuring that security postures are capable of withstanding current and emergent threats. This not only helps in identifying potential breaches but also assists in fortifying the organization's defenses against similar future attacks. 

\begin{figure}[!b]
    \centering
    \includegraphics[width=\linewidth]{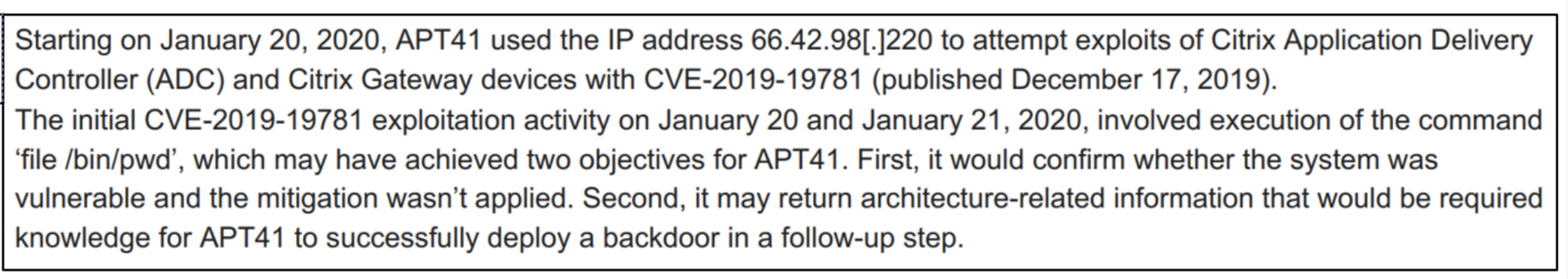}
    \caption{Mandiant's APT41 threat report~\cite{googleblog2024}.}
    \label{fig:report_examples}
\end{figure}

\subhead{Problem Statement}
The extraction of attack testflow from report 
is a critical but time-intensive task requiring extensive cybersecurity knowledge and analytical skills. Currently, the process is predominantly manual~\cite{wiliam2024interview}, leading to inefficiencies and potential for oversight. This is particularly problematic when considering the need for a comprehensive understanding of different APT campaigns, where integrating insights from multiple reports is essential.
{\ahmadou Fig.~\ref{fig:report_examples} presents an excerpt of APT41 reports from Mandiant published in 2024~\cite{googleblog2024}, 
where security experts need to identify specific elements involved in the attacks.}
%

%
Automating the extraction of testflow from diverse and unstructured threat reports would dramatically streamline this process, reduce time and resource expenditure, and increase the accuracy and comprehensiveness of the testflow. 
{\ahmadou APT attacks are very sophisticated, and extracting the right testflow from the reports could be cumbersome and error-prone~\cite{alshamrani2019survey},} resulting in incomplete, incorrect, or misinterpreted information. 
The process requires a lot of time and human resources, which can be costly and inefficient in the long run. 

%
To address these concerns, it is important to consider how to automate the extraction of attack steps and generate testflows corresponding to attack vectors to improve the effectiveness of the incident response.
{\faissal Our approach does not focus on high-level testflow based on techniques (\ie TTPs) instead, we prioritize low-level testflow that captures the granular details of an attack to provide more comprehensive information and a more detailed testflow.}

\section{Related Work}
\label{sec:relatedwork}

\titlespacing{\subsection}{0pt}{*0.2}{*0.2}

\subhead{Cyber Threat Intelligence (CTI)}
CTI plays a crucial role in cybersecurity by enabling organizations to anticipate, prevent, and respond to cyber threats through structured intelligence sharing.
Recent advancements focus on dynamic intelligence extraction, mapping attack patterns to frameworks (\eg MITRE ATT\&CK), and leveraging machine learning for automation. However, existing CTI frameworks primarily address knowledge representation rather than attack testflow generation. While prior work (\eg ~\cite{10.1145/3607199.3607208}) has explored CTI sharing challenges, frameworks for structured intelligence extraction remain a gap in integrating language models for real-time, automated cybersecurity response.
While threat reports compile insights into detailed narratives tailored for security practitioners, the increasing complexity and volume of these reports have outpaced the capabilities of traditional approaches, requiring the use of new efficient methods. 
\subhead{Language Models in Cybersecurity}
LMs (\eg BERT~\cite{devlin2018bert}, RoBERTa~\cite{liu2019roberta}, MUM~\cite{nayak2021mum}) have been widely explored in cybersecurity for tasks such as intrusion detection, malware analysis, phishing detection, and vulnerability assessment~\cite{motlagh2024largelanguagemodelscybersecurity}. While these studies demonstrate the potential of language models in various security applications, they primarily focus on general cybersecurity challenges rather than the automation of attack testflow extraction from CTI reports. Despite advances in generative AI for cybersecurity~\cite{bhusal2024securebenchmarkinggenerativelarge}, there remains a gap in using them for attack testflow generation.

\subhead{Attack Testflow Extraction}
{\sepehr 
A testflow represents the sequence of attack steps, allowing organizations to assess vulnerabilities and validate security measures~\cite{guo2023framework}.
Several approaches exist for extracting attack testflows such as, information extraction~\cite{9946567}, ontology-based Methods~\cite{10266600}, and Machine Learning-based extraction~\cite{9526808}.
%
{\ahmadou Building on these foundational approaches, several systems have been proposed to further automate this process.} EXTRACTOR~\cite{satvat2021extractor} formalizes attack steps into evidence graphs but relies heavily on NLP accuracy, limiting its robustness for unstructured CTI. IC-SECURE~\cite{kremer2023ic} facilitates playbook generation but does not generalize to real-time attack testflow extraction. Similarly, THREATRAPTOR~\cite{gao2021enabling} and TTPDrill~\cite{husari2017ttpdrill} focus on IoC extraction and ATT\&CK mapping, yet their rule-based techniques lack adaptability to evolving threats, while LADDER~\cite{10.1145/3607199.3607208} aims to extract attack patterns from CTI and map them to the MITRE ATT\&CK framework.
Despite these advancements, automated attack testflow extraction from unstructured reports remains an underexplored area. Current methods focus on structured knowledge representation rather than real-time adaptive testflow generation. Addressing this gap would enhance security operations by transforming unstructured knowledge from CTI into actionable intelligence for proactive defense.}
 
{\revision
Our approach, \sol, systematically combines domain-trained language models (\eg BERT) for contextual analysis with sequence analysis to extract structured testflows directly from unstructured threat reports. \sol focuses not only on identifying entities and actions but also on reconstructing their order and contextual meaning to generate relevant testflows. This allows \sol to generate testflows with semantic coherence and operational relevance.
}

\section{Automated Attack Testflow Extraction}
\label{sec:sol}

\subsection{System Overview}
%

\begin{figure*}[!t]
    \centering
    \includegraphics[width=1\linewidth]{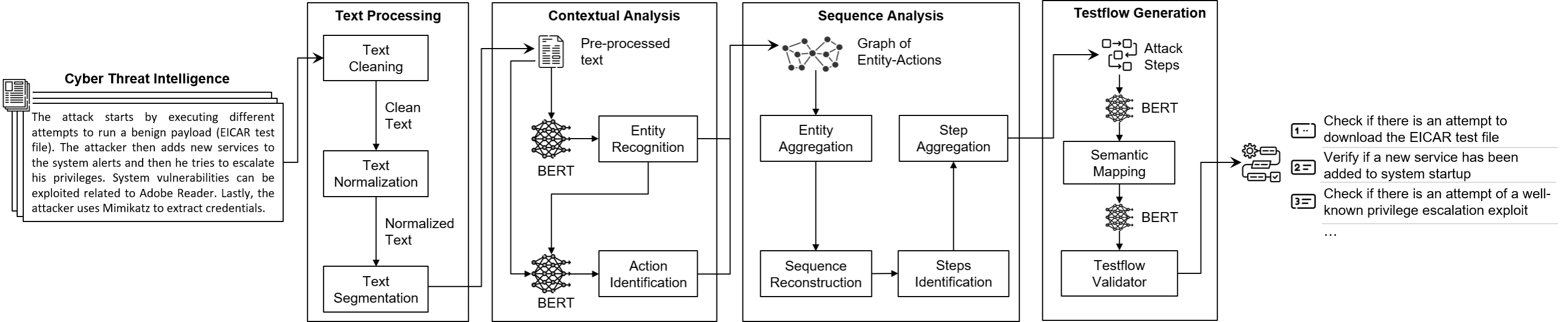}
    \caption{High-level overview of \sol.}
    \label{fig:sol}
    \vspace{-4mm}
\end{figure*}

\subhead{Overview}
\sol, depicted in~Fig.~\ref{fig:sol}, has four main steps: 
\begin{enumerate*}[(1)]
    \item text pre-processing: involve standardizing the text to be ready for testflow extraction. 
    \item contextual analysis: interpret its context using semantics and word relationships. Language model (\eg BERT) is used to extract relevant information, detect subtexts, and identify key elements such as named entities and action verbs that are crucial for generating relevant testflows.
    \item sequence analysis: the identified entities and their context are analyzed to reconstruct the sequence and logic of the attack steps, providing a clearer understanding of the attack pattern and assisting in the testflow generation.
    \item The final step is testflow generation. Using insights from the contextual analysis, we automatically build testflows.  
    {\ahmadou These are generated using LM guided by rules trained on the analyzed data, resulting in a concrete list of testflows for the APT attack sequences listed in the report. To improve this process, we further include a testflow validator to filter out invalid or redundant testflows and refine those that can be adjusted to meet validity criteria.}
\end{enumerate*}

\subhead{Novelty}
The core solution of \sol involves three novel steps:
\begin{enumerate*}[(1)]
    \item \textit{Threat Contextualization}: refers to analyzing and understanding security events, incidents, or threats in the report. {\ahmadou To define the security context, we include how a particular vulnerability or threat affects an organization in terms of its network configurations, software usage, business processes, and protective measures.}
    \item \textit{Attack Steps Analysis}: refers to the systematic examination of all activities performed by the threat actor to achieve their objectives. The goal is to identify the various stages through which attackers move from the initial point of breaching security measures to their ultimate goal, such as stealing data, compromising systems, or achieving persistent presence.
    {\ahmadou This method goes beyond existing approaches by offering a complete, contextualized, and actionable understanding of an adversary's movements.}
    \item \textit{Automated Testflow Extraction}: refers to the process of using {\faissal extracted information} in combination with various {\faissal techniques} to generate testflows.
    {\ahmadou Current methods often generate generic testflow, lacking relevance to specific cybersecurity scenarios. \sol uses cybersecurity-specific data and techniques to ensure that the generated testflows are directly aligned with the described attack.}
\end{enumerate*}

\smallskip
\subhead{Design Challenges}
We present below the main challenges and how we address each.
\begin{enumerate*}[(i)]
    \item \textit{Complexity of Cybersecurity Language}: the complexity involved in using specific terms related to cyber protection. 
    In addition, this language often requires deep knowledge of technical concepts, making it difficult for NLP efforts.
    \item \textit{Data Scarcity and {\ahmadou Low-Quality reports}}: {\ahmadou
    insufficient relevant data poses a significant challenge in training machine learning models and conducting meaningful analysis. 
    Additionally, reports with low-quality, incomplete, or complex language further complicate the analysis process. Standard language models, 
    struggle to handle such reports, failing to generate meaningful outputs.}
    \item \textit{Extraction of Relevant Information}: 
    {\ahmadou general purpose NLP models, 
    are designed for broad language tasks. However, they lack specialized capabilities for cybersecurity contexts.
    }
\end{enumerate*}
In our work, we use LM and NLP to perform contextual analysis to extract cybersecurity entities and actions. {\ahmadou \sol integrates spaCy and fine-tunes them with BERT to enable accurate recognition of all the relevant entities in the cybersecurity domain.} 
We also use sequence analysis to understand the context and relationships between entities, which helps in extracting testflows. Finally, we use semantic mapping of these sequences into a related test.

\subsection{Text Processing}
This phase involves several NLP techniques~\cite{10128641} to refine raw, unstructured data such as: 
a) text cleaning, which prepares raw text data for further processing and analysis, b) normalization, which 
converts data into a consistent and standard format, simplifies analysis 
and c) segmentation, which simplifies and enhances the effectiveness of data processing.
The result of the text processing is a pre-processed text, which has been both cleaned and structured. This prepared text serves as the input for the contextual analysis, enabling more sophisticated analytical processes.

\subsection{Contextual Analysis}
\label{context_analysis}
Extracting and interpreting the meaning of text requires considering its cybersecurity context, whether within a sentence, paragraph, or broader document. Contextual analysis is vital for understanding both the content and implications of the text. 
\sol employs LM to extract key threat-related entities and action verbs within the threat reports.
%

\smallskip
\subhead{\sol NER Approach}
{\ahmadou Traditional NER often struggles to recognize and extract emerging cybersecurity entities~\cite{10428218} such as threat actors, malware names, attack types, and techniques, which are essential for generating meaningful testflows.
\sol addresses this limitation by: 
\begin{enumerate*}[(i)]
    \item Pre-training models using general datasets like spacy.
    \item Fine-tuning BERT on our custom cybersecurity dataset, enabling the extraction of cybersecurity-specific entities.
\end{enumerate*}
}

\smallskip
\subhead{LM Training and Fine-Tuning}
To train models that mimic the expertise of security experts, a security dataset is required for LM training.
\sol leverages  
BERT~\cite{devlin2018bert} to analyze and extract security information from threat reports, enabling precise identification of hidden threats.
BERT's advanced contextual comprehension allows \sol to detect subtle linguistic nuances in cyber threat reports. By utilizing BERT's pre-trained language representations, we enhance the accuracy and efficiency of our threat intelligence and analysis. 
{\revision
For cybersecurity applications, BERT is fine-tuned on labeled NER datasets to extract domain-specific entities from CTI reports with high precision. Unlike large generative models (\eg ChatGPT, LLaMA), which are prone to hallucinations, producing plausible but incorrect or unverifiable information. BERT offers greater stability and reliability. Since BERT is designed for discriminative tasks rather than text generation, it avoids the unpredictability of generative LLMs and produces consistent outputs. This makes BERT a strong and justified choice for structured information extraction in the cybersecurity context.
Additionally, \sol remains compatible with other LM-based models, ensuring adaptability across various industries and applications.}

\begin{figure}[!b]
  \includegraphics[width=.9\linewidth]{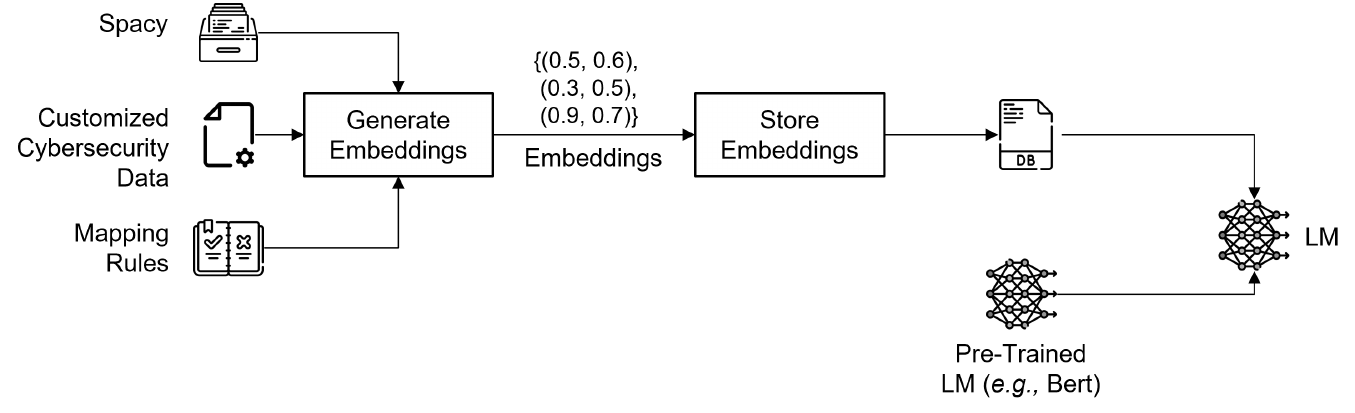}
  \caption{\small Overview of steps to train/fine-tune the BERT model.}
  \label{fig:finetune}
\end{figure}

During the fine-tuning (see Fig.~\ref{fig:finetune}), we used 
SpaCy, an NLP model specifically designed for named entity recognition (NER). However, despite being tailored for NER tasks, the model struggled to accurately identify certain entities.
To overcome this, we fine-tuned the model with a customized dataset (see subsection~\ref{sec:dataset}), significantly enhancing our entity recognition accuracy.
\noindent {\sepehr The customized dataset refines cybersecurity entity categorization into eight key groups to better capture underrepresented concepts, such as:
\begin{enumerate*}[(i)]
    \item {\tt TEC} (Technique): methods or strategies employed by threat actors to exploit vulnerabilities or execute malicious actions, \eg~{\tt phishing}.
    \item {\tt TAC} (Tactic): attack strategies from MITRE ATT\&CK, \eg~{\it initial access}.
    \item {\tt THA} (Threat Actor): known adversarial groups like {\it APT41}.
\end{enumerate*}}
%
%
Additionally, we also trained the model to map attack steps into the testflow using a set of predefined mapping rules (see subsection \ref{sec:saemanticmapping}). The objective is to efficiently correlate the analyzed text with specific attack descriptions, leveraging the accurately identified entities from the previous step.

\smallskip
\subhead{Cybersecurity Action Identification}
This process focuses on finding action verbs within sentences, and helps uncover behavior patterns, automate analysis, and understand the relationships between entities. Action identification involves pinpointing verbs and verb phrases, such as \textit{``downloading file'', ``adding new services'', ``exploiting vulnerabilities''}, and understanding how these actions related to entities in the context -- which is very challenging using traditional NLP. 
%
Leveraging LMs for action verb extraction enhances the analysis of threat reports by identifying relevant actions based on the used language and learned patterns.
%

\subsection{Sequence Analysis}
This involves examining text to extract valuable information. 
A sequence is an ordered list of elements or events following a specific pattern, whether temporal or sequential. Analyzing the identified NER entities and action verbs within their context helps reconstruct the sequence and logic of attack steps. 
Advanced techniques focus on entity aggregation, sequence reconstruction, step identification, and step aggregation, each playing a distinct role in interpreting structured data.

\smallskip
\subhead{Entity Aggregation}
Entity aggregation combines related entities based on their occurrence and context within the text by grouping them into a single aggregated entity. This process reduces redundancy, enhances analysis efficiency, and simplifies the management of entities. The primary goal is to minimize the number of entities for easier analysis. 
     Using the extracted NER and identified action verbs, we can aggregate entities based on their types and associated actions. 
    %
For example, ``{\it The attacker}'' and the pronoun ``{\it he}'' are recognized as the same entity, despite the shift in reference. 

\smallskip
\subhead{Sequence Reconstruction}
Sequence reconstruction involves assembling segments into a coherent order in order to arrange attack steps based on their narrative flow or temporal cues like ``{\it first}'', ``{\it then}'', or ``{\it finally}''. 
{\ahmadou The narrative flow from the text provides the order of operations that the attacker followed, and we map each action to an appropriate phase and tactic based on MITRE ATT\&CK framework.}
{\faissal To achieve sequence reconstruction, we rely on a natural process that mimics how humans understand the flow of events by interpreting their relationships and context.}

\smallskip
\subhead{Step Identification}
This step involves detecting and parsing distinct actions or events within a sequence, highlighting each step's characteristics and role. This step clarifies the sequence by identifying relationships between steps, such as how initial access through a \textit{phishing email} leads to \textit{malware installation}. This process is essential for tracing the flow of activities, enabling a deeper understanding of attack schemes, patterns, or workflows within complex sequences.
To accomplish this task, \sol 
    firstly, analyzes the relationships between the identified NER and the corresponding action verbs, as well as labeled with step identifiers (\eg~{\it Step 1}, {\it Step 2}); and then
    orders the steps using natural step and logical flow based on temporal sequence indicators such as ``{\it First}'', ``{\it Next}'', ``{\it Then}'', and ``{\it Finally}''.
%
{\faissal In this process, the relationships between NER and actions guide the step ordering.
This relationship is what drives the logical flow of attack, helping to determine which step comes before or after another.}
%


\smallskip
\subhead{Step Aggregation}
This step is similar to entity aggregation but focuses on action verbs. Texts may describe similar steps differently or repeat them, so consolidating these avoids redundancy and emphasizes unique attack vectors.
One key method is combining identified steps into broader stages or phases, making the sequence more concise and easier to understand. Grouping smaller actions into strategic categories simplifies the attack or testflow analysis. For this, we use a clustering algorithm to group related steps based on feature similarity.

In doing so, we follow the following process:
\begin{enumerate*}[(i)]
    \item \textit{Grouping}: When multiple steps involve the same entity, they can be grouped into one single aggregated step.
 
    \item {\faissal \textit{Labeling}: we label each aggregated step with a phase name according to MITRE ATT\&CK framework to each aggregated step. 
    {\ahmadou Rule-based matching (defined based on tactics in MITRE ATT\&CK) is used to match specific actions or keywords to the corresponding tactic. 
    }
 }
    \item {\faissal \textit{Validation}: we ensure that the steps are logically grouped and represent meaningful phases of the attack. The validation consists of comparing the steps and aggregated phases against the MITRE tactics and techniques to ensure that all relevant stages of the attack are covered.}
\end{enumerate*}

\subsection{Testflow Generation}
Testflow generation involves creating detailed tests or sequences to verify whether the described attack exists within a system, ensuring security measures are effective. 
%
%

\smallskip
\subhead{Semantic Mapping}
\label{sec:saemanticmapping}
To enhance semantic understanding, context-based action verbs are linked directly to the identified entities. We created a set of labeling rules for generating testflow, detailing how entities and actions should be structured. 
These rules are used to train the BERT model, enabling it to automatically generate detailed, contextually appropriate testflows. 
%

\smallskip
\noindent \textit{Mapping rules}: 
BERT is meticulously trained using a set of mapping rules that convert detailed security incident descriptions into specific, actionable testflow instructions
(see Fig.~\ref{fig:finetune}).
These rules help verify the presence and impact of various cybersecurity threats. Each rule is linked to a particular security issue (\eg malware, ransomware) and defines the corresponding action (\eg run, encrypt) along with a related test command.
{\ahmadou These mapping rules are integral to the model's ability to effectively verify the presence and assess the impact of various cybersecurity threats within a system. Each rule is carefully linked to a specific security issue, such as malware infections, ransomware attacks, or unauthorized access attempts. The rules define the corresponding actions required to address these issues, such as executing a script, encrypting data, or initiating a network scan. Furthermore, these rules are not arbitrarily chosen; they are carefully selected and designed based on the core primitives of testflow, ensuring that they align with fundamental security operations.  By grounding the rules in these core primitives, we ensure that the testflows generated by the LM model are not only relevant but also comprehensive in covering the essential aspects of testflow generation.}
%
    


\smallskip
{\ahmadou
\subhead{Testflow Validator}
To improve the quality and reliability of our testflow generation process, we introduce a tesflow validator component. This component serves two main purposes:
\begin{enumerate*}[(a)]
    \item \textit{filtering out invalid tests:} ensure that testflow generated by \sol are valid and meet the required criteria, and
    \item \textit{eliminating redundancy:} remove duplicate or redundant testflow to streamline the output.
\end{enumerate*}
The output of this component is manually cross-validated to confirm the effectiveness of the filtering and refinement process. During this manual iteration, any remaining redundant or invalid testflows are identified and removed. 
}





    

\begin{figure}[!h]
        \centering
        \includegraphics[width=.8\linewidth]{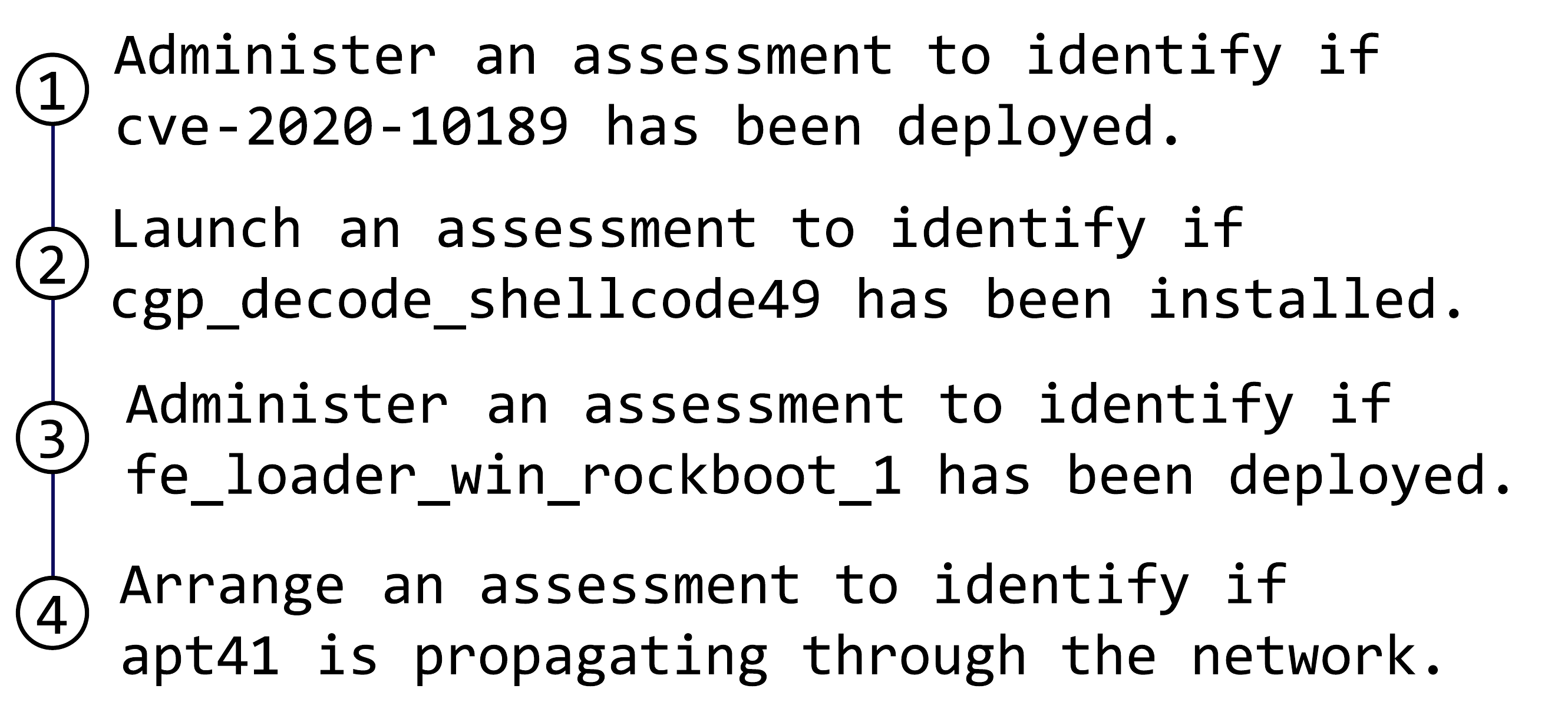}
        \caption{Example of testflows generated by \sol.}
        \label{fig:testflow_example}
\end{figure}

Fig.~\ref{fig:testflow_example} shows examples of testflows generated by \sol.
The generated testflows align with standard security operations and highlight how \sol effectively proposes relevant testflow for the used threat report. 

\section{Implementation}
\label{sec:implementation}
\subsection{Environment}
We implemented \sol using Python 3 programming language. Our experiments were conducted on Intel Xeon E312xx, 6 vCPU @ 2.69Ghz, with 32GB RAM running Ubuntu 20.04. For linguistic analysis, we employed the \texttt{spaCy}\footnote{spaCy: https://spacy.io/} library, a comprehensive tool for natural language processing in Python, to extract action verbs. Specifically, we utilized the {\tt en\_core\_web\_sm} model.

\subsection{Datasets}
\label{sec:dataset}
\subhead{Threat Reports}
We use threat reports from the APTNotes repository\footnote{APTNotes repository: \url{https://github.com/aptnotes/data/}}, which gathers different threat intelligence from vendors like Mandiant and 
TrendMicro.
Given the lack of comprehensive testflows in existing reports and the difficulty of manual extraction, we focus solely on APT41 for the experiment discussion, a sophisticated APT with diverse techniques and multiple public reports. 
%
%
%
Indeed, we use five different APT41 reports published by different threat intelligence companies describing different APT41 campaigns. 
For example, the Mandiant \#1~\cite{mandiant2024} is 68 pages long and covers backdoors, reconnaissance, and credential theft in some detail. This detail allows for the extraction of highly detailed and valid testflows, reducing redundancy.
Other reports, such as Mandiant \#2~\cite{googleblog2023} and Mandiant \#4~\cite{googleblogAPT41states} are shorter 9 pages each and often do not have the level of granularity required to produce comprehensive testflows. Consequently, the extracted testflows may be more general and tend to be redundant because of the small amount of usable detail.
%

\textit{Quality of threat reports:}
The quality and structure of the reports play a critical role in the reliability of the testflows that are extracted from them.
BertScore~\cite{saadany2021bleu} is a metric that uses deep contextual embeddings from pre-trained models like BERT to evaluate semantic similarity. In contrast to traditional metrics that focus on syntactic matches, BERTScore captures nuanced relationships between words and phrases, taking into account synonyms, paraphrases, and contextual shifts. 
This makes it an ideal tool for assessing the semantic similarity of texts. We use BERTScore to assess the semantic similarity of reports processed by our solution, ensuring meaningful and coherent outputs. Additionally, we compare testflows generated by our system with those manually created by practitioners to validate their alignment with domain expertise.

\begin{figure}[!b]
    \centering
    \includegraphics[width=.9\linewidth]{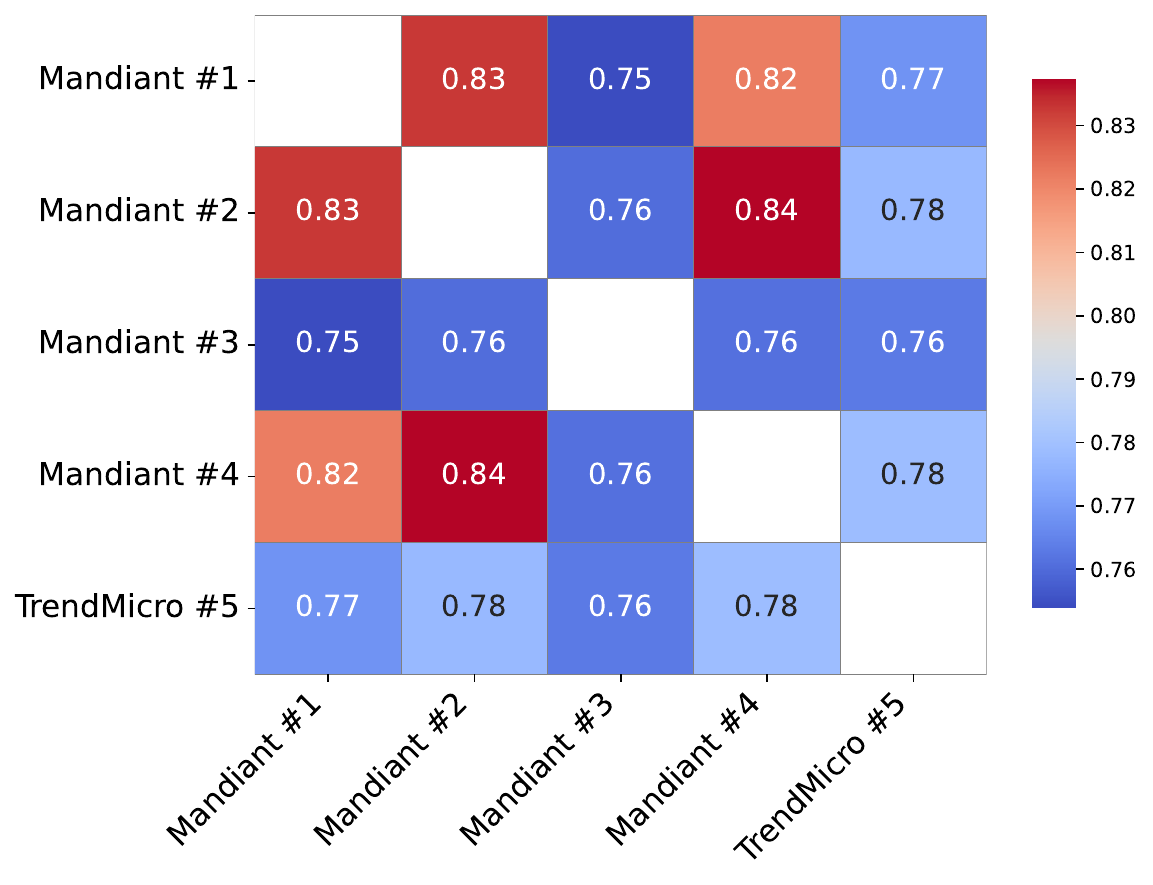}
    \caption{Semantic Similarity between reports using BertScore~\cite{saadany2021bleu}.}
    \label{fig:similarity_reports}
\end{figure}

Fig.~\ref{fig:similarity_reports} shows the BertScore of the used threat report. We can see that reports Mandiant~\#2 and Mandiant~\#4 have the highest similarity (0.84), indicating shared patterns in their testflows. In contrast, Mandiant~\#3 exhibits the lowest similarity, particularly with Mandiant~\#1 and Mandiant~\#4, suggesting distinct attack scenarios (see Section~\ref{subsec:sim_testflow} for further analysis).
Indeed, the report quality significantly impacts generated testflows. Well-structured, detailed reports improve accuracy and relevance, while similar reports yield consistent testflows, reinforcing the importance of high-quality input data.

\subhead{Testflow Primitives}
In this work, we focused on the core functions of attack steps and derived 30 core cybersecurity primitives through NERs. Using specific mapping rules, NER entities were linked to action verbs in security testing scenarios. 
We also developed 44 mapping rules 
to structure testflows based on the identified NER and action verbs. These fine-tuned primitives target various security issues, enhancing the system's ability to identify and mitigate vulnerabilities effectively.

%

\subsection{Benchmarked Models}
\label{sec:other_Models}
In the absence of established tools for generating testflows from threat reports, we introduce four distinct models to benchmark and compare their performance against \sol's results. These models leverage the capabilities of {\tt ChatGPT-4o-mini}\footnote{GPT-4o mini: \url{https://platform.openai.com/docs/models/gpt-4o-mini}} and {\tt Llama3-70b}\footnote{Llama3: \url{https://ai.meta.com/blog/meta-llama-3/}}  as foundational components:

\begin{itemize}
    \item Off-the-shelf Models: {\tt ChatGPT-off-the-shelf} and {\tt Llama-off-the-shelf} utilize the raw power of the underlying LLMs without any modifications. In these configurations, the threat report is directly fed into the LLM, and the outputted testflows are analyzed to gauge the baseline performance.
    
    \item {\ahmadou Off-the-shelf Models with Partial \sol's Pipeline: {we integrated the pipeline of \sol (PFG), as depicted in Fig.~\ref{fig:sol}, into Off-the-shelf models                
   {\tt ChatGPT-with-PFG} and {\tt Llama-with-PFG}} to demonstrate the impact of \sol's architecture on testflow generation, providing a clear evaluation of our system's design efficacy.} 

\end{itemize}
\subsection{Ground truth data}
In the absence of a public testflow dataset, we enlisted the expertise of two security practitioners to manually create testflows for APT41's threat report. Table~\ref{tab:manual_generated_testflow} shows the number of manually generated testflow (\ie ground truth) of both security practitioners per each report. 
We also verified the combined testflows to eliminate any redundant or duplicate testflows, and then applied a cross-validation by both security practitioners to ensure the accuracy of our results. The unique testflows are used as a ground truth.
To simulate real-world scenarios, we instructed one security practitioner to focus on creating high-level and generalized testflows, while the other security practitioner concentrated on developing more granular and specific testflows. This dual approach mirrors the diversity in practices among security practitioners. Seasoned practitioners often prioritize fewer and more targeted tests, whereas less experienced practitioners may employ a larger number of tests to ensure comprehensive coverage.

\begin{table}[!ht]
    \centering
    \caption{Summary manual generated testflow and their characteristics (\ie redundant and unique).}
    \label{tab:manual_generated_testflow}
    \resizebox{\linewidth}{!}{%
    \begin{tabular}{rc c c c c}
        \toprule
        \textbf{Threat Report} & Mandiant \#1 & Mandiant \#2 & Mandiant \#3 & Mandiant \#4 & TrendMicro \#5 \\ \midrule
        \textbf{Manual Testflows} & 27 & 16 & 42 & 48 & 33 \\ 
        \textbf{Redundant Testflow} & 3 & 5 & 9 & 6 & 8 \\ 
        \textbf{Unique Testflow} & 24 & 11 & 33 & 42 & 25 \\
        \bottomrule
    \end{tabular}
    }
\end{table}


   


%


\begin{figure*}[!ht]
    \centering
    \subfigure[Mandiant \#1~\cite{mandiant2024}]{ 
        \label{fig:Mandiant_1-2019-08-07}
        \includegraphics[width=.3\linewidth]{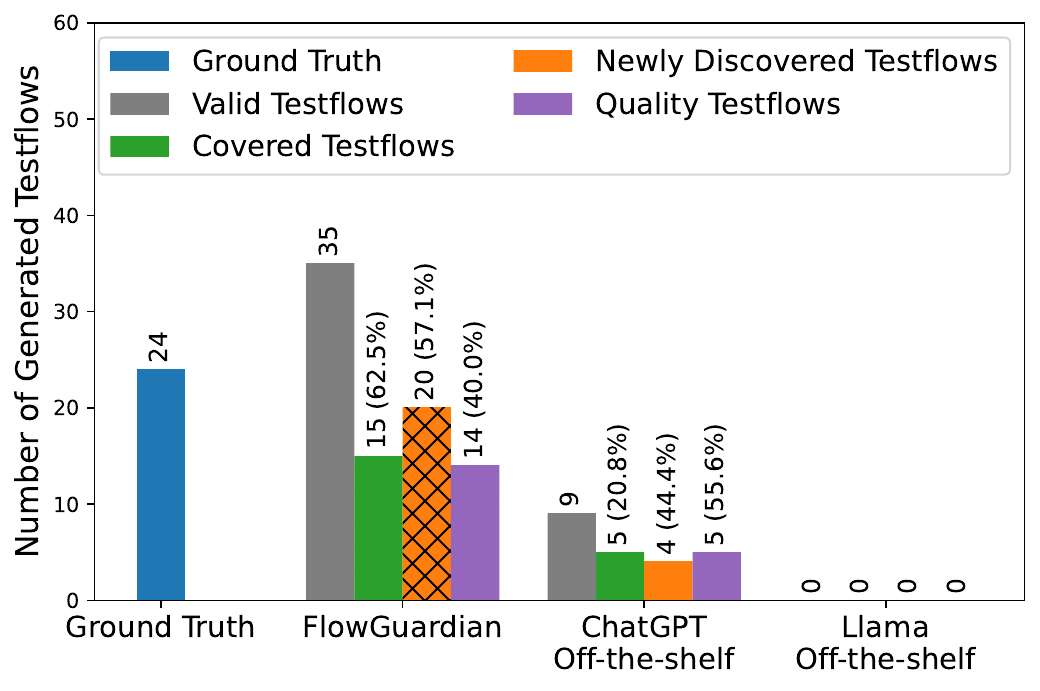}
    }
    \subfigure[Mandiant \#2~\cite{googleblog2023}]{ 
        \label{fig:Mandiant_2_2019-08-19}
        \includegraphics[width=.3\linewidth]{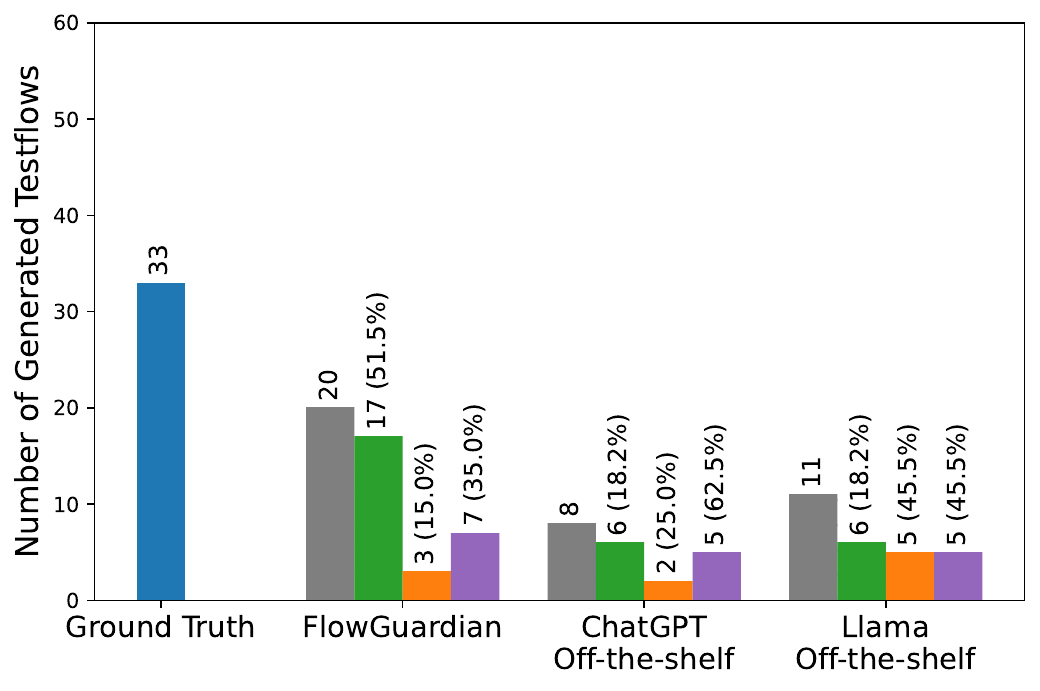}
    }
    \subfigure[Mandiant \#3~\cite{googleblog2024}]{ 
        \label{fig:Mandiant_3_2020-03-25} 
        \includegraphics[width=.3\linewidth]{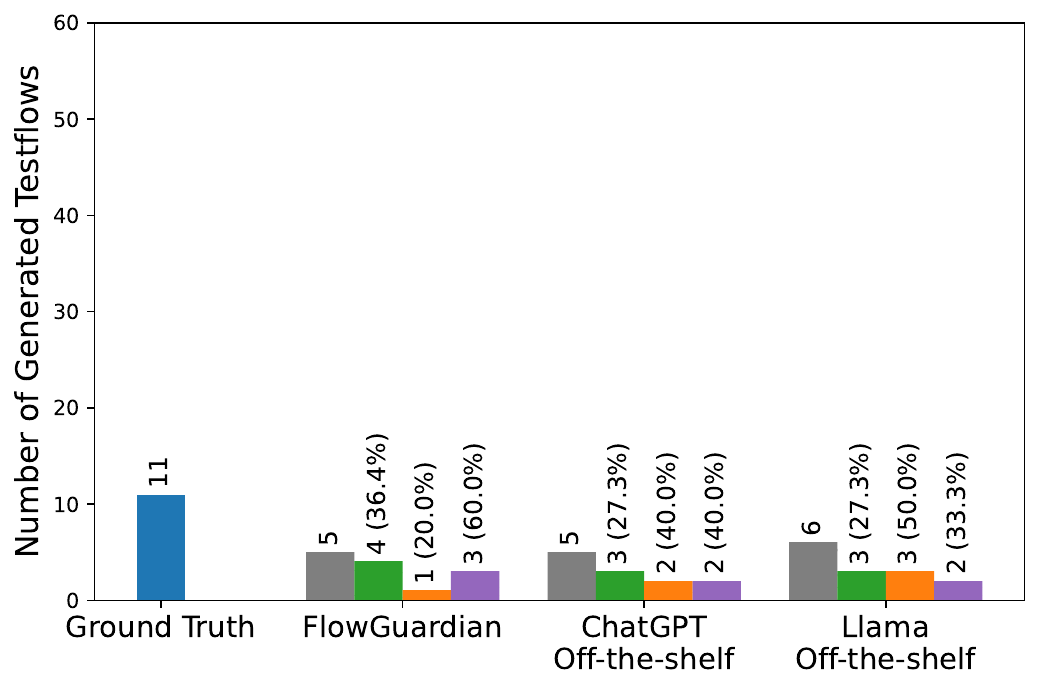}
    }
    \subfigure[Mandiant \#4~\cite{googleblogAPT41states}]{ 
        \label{fig:Mandiant_4_2022-03-08}
        \includegraphics[width=.3\linewidth]{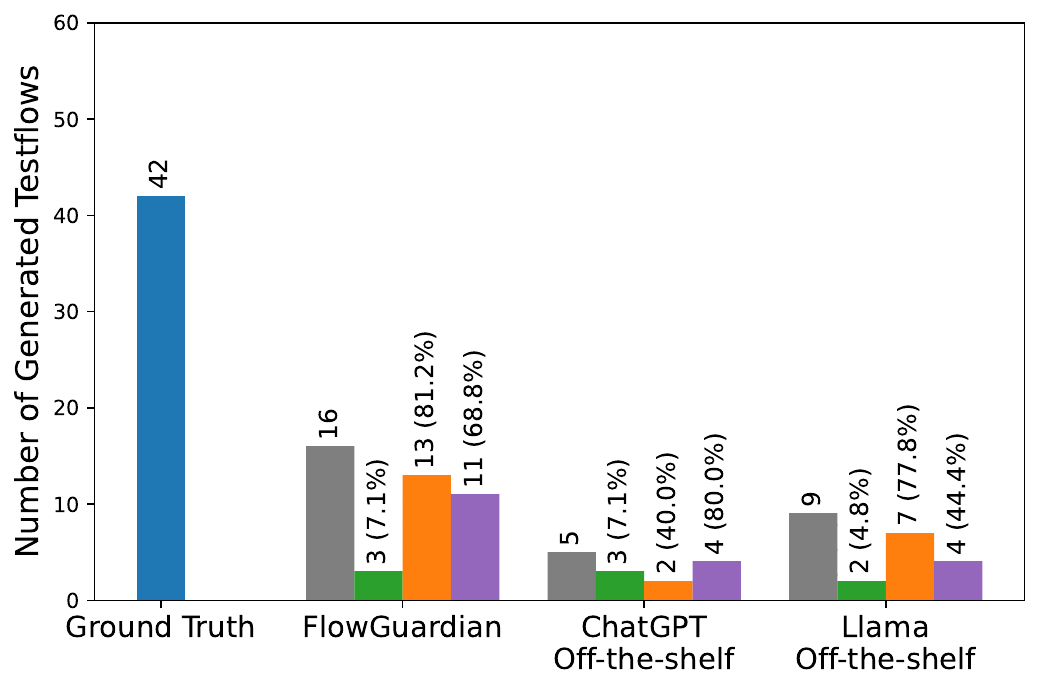}
    }
    \subfigure[TrendMicro \#5~\cite{trendmicro2021earth}]{ 
        \label{fig:TrendMicro_2021-08-24}
        \includegraphics[width=.3\linewidth]{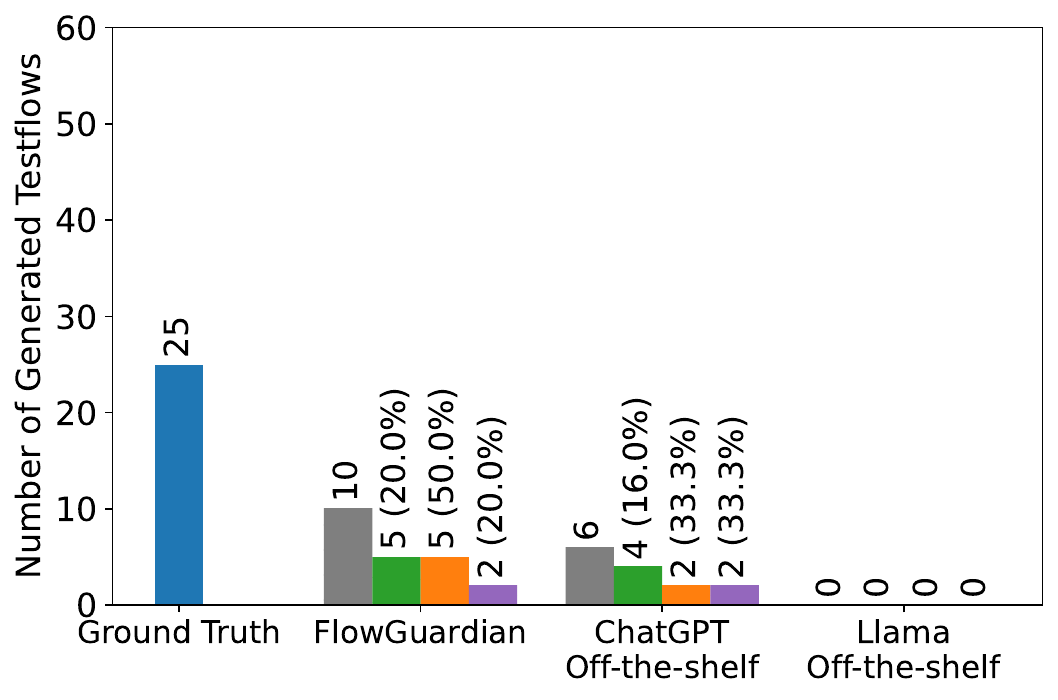}
    }   
    \caption{ \ahmadou Comparison of testflows generated by \sol and Off-the-Shelf Models across five APT41 reports.}
    \label{fig:fg_against_off_the_shelf}
    \vspace{-4mm}
\end{figure*}

\section{Evaluation}
\label{sec:evaluation}

\subsection{Evaluated Metrics} 
To evaluate the testflows generated either by an automated solution or a security practitioner, we introduce four metrics: \textit{testflow coverage}, \textit{testflow validity}, \textit{newly discovered testflow}, and \textit{quality testflow}.

\smallskip
\subhead{Testflow Coverage}
The testflow coverage is defined as the proportion of manually generated testflows that are successfully captured by the testflows produced by a given model. A valid testflow exact match or through semantic equivalence with the manually generated testflow. A semantic equivalence means that more than one testflow will collectively match one testflow.

\smallskip
\subhead{Testflow Validity}
The testflow validity refers to testflow that aligns with the practical requirements and expectations of security practitioners, and satisfies both of the following conditions:
\begin{enumerate*}[(i)]
    \item \textit{relevance}: testflow is relevant and makes sense from the security perspective, reflecting real-world scenarios;
    \item \textit{applicability}: testflow is applicable in the context of testing APT in question.
\end{enumerate*}

\smallskip
\subhead{Newly Discovered Testflow}
The newly discovered testflow refers to a testflow generated by an automated solution that is valid yet not part of the testflows originally created by practitioners. These testflows are cross-validated by practitioners to confirm their validity and ensure they represent new, previously uncovered tests that were not addressed in the practitioners' original test set.

\smallskip
\subhead{Quality Testflow}
A quality testflow refers to a testflow that is highly specific to the APT41 as described in the report. Among the testflow generated by the automated solution, some may be generic (applicable to various scenarios), while others are directly linked to the APT41 TTPs. The focus is on capturing the unique behaviors of the threat actor to ensure the testflow relevance.

{\ahmadou
\subsection{Performance of \sol vs Off-the-Shelf Models}
\label{subsec:Experiment1}



\begin{figure*}[!ht]
    \centering
    \subfigure[Mandiant \#1~\cite{mandiant2024}]{
        \label{fig:Mandiant_1_with_PFG}
        \includegraphics[width=.3\linewidth]{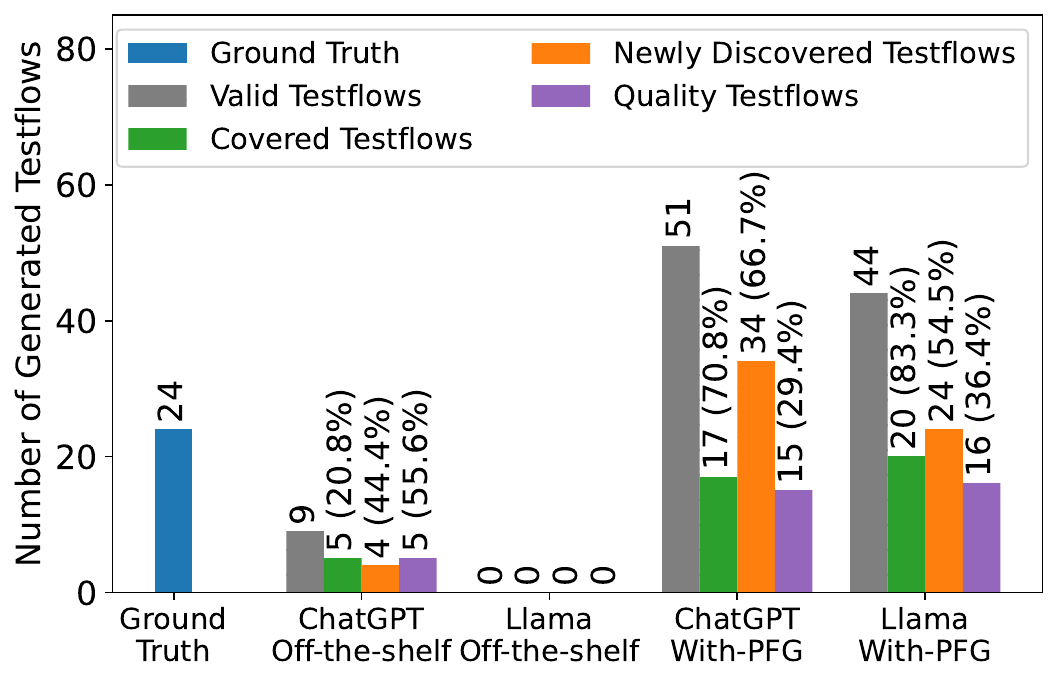}
    }
    \subfigure[Mandiant \#2~\cite{googleblog2023}]{
        \label{fig:Mandiant_2_with_PFG}
        \includegraphics[width=.3\linewidth]{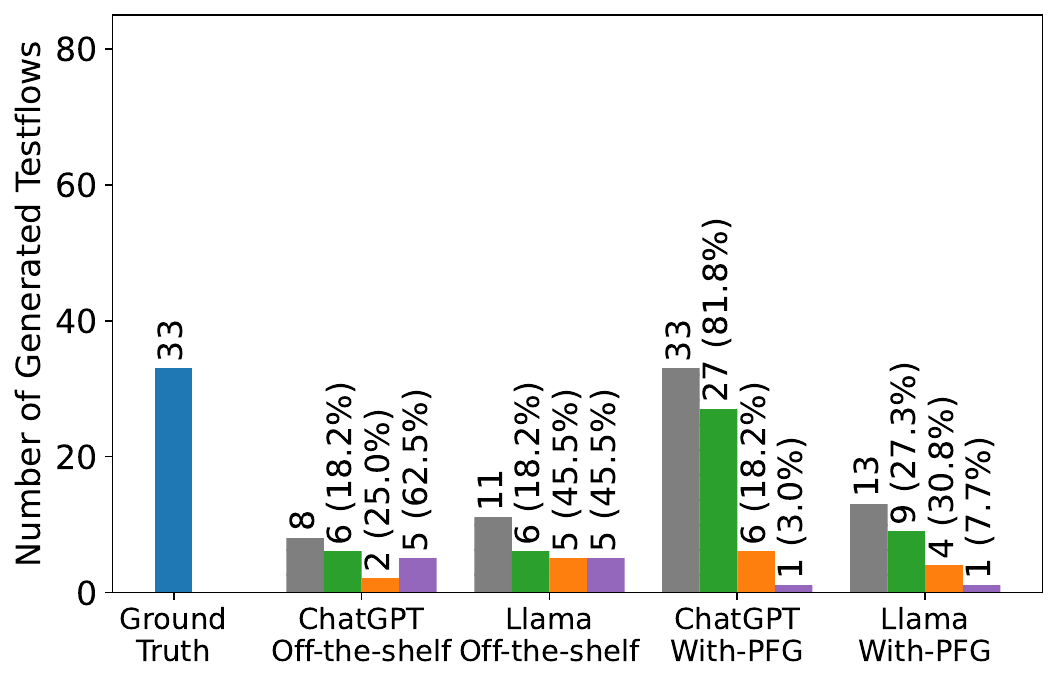}
    }
    \subfigure[Mandiant \#3~\cite{googleblog2024}]{    \label{fig:Mandiant_3_with_PFG}
        \includegraphics[width=.3\linewidth]{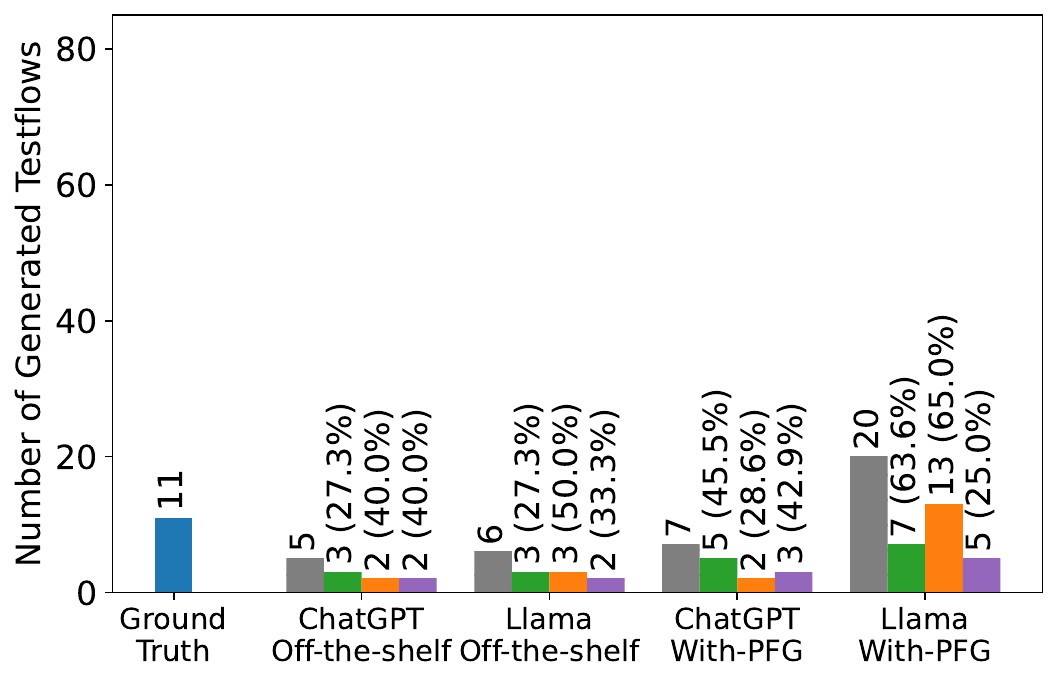}
    }
    \subfigure[Mandiant \#4~\cite{googleblogAPT41states}]{
        \label{fig:Mandiant_4_with_PFG}
        \includegraphics[width=.3\linewidth]{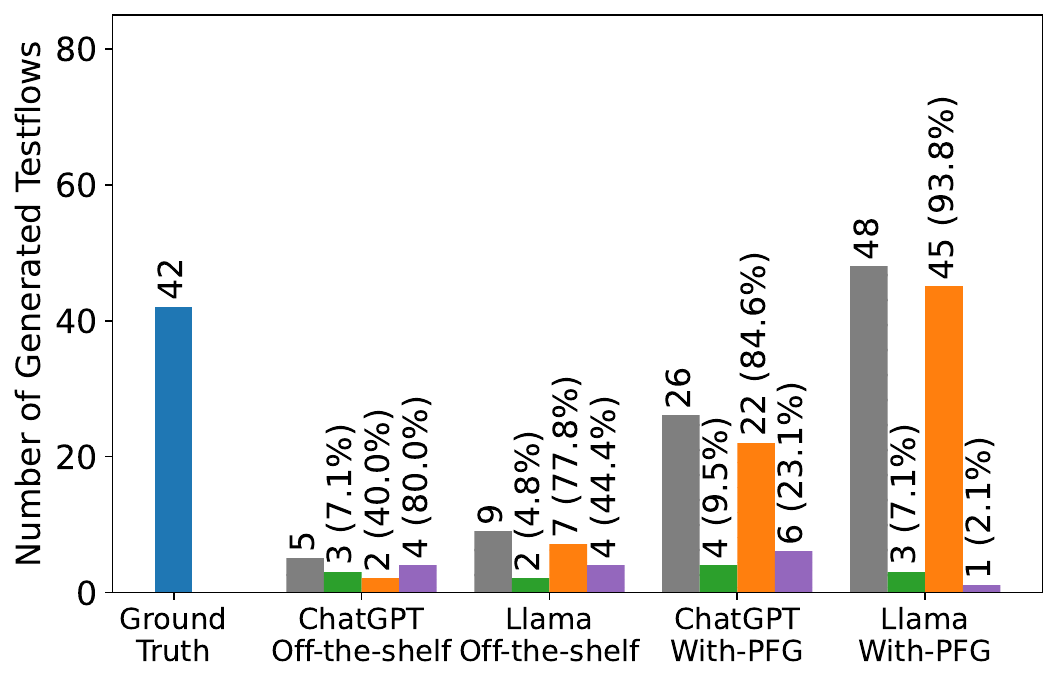}
    }    
    \subfigure[TrendMicro \#5~\cite{trendmicro2021earth}]{
        \label{fig:TrendMicro_with_PFG}
        \includegraphics[width=.3\linewidth]{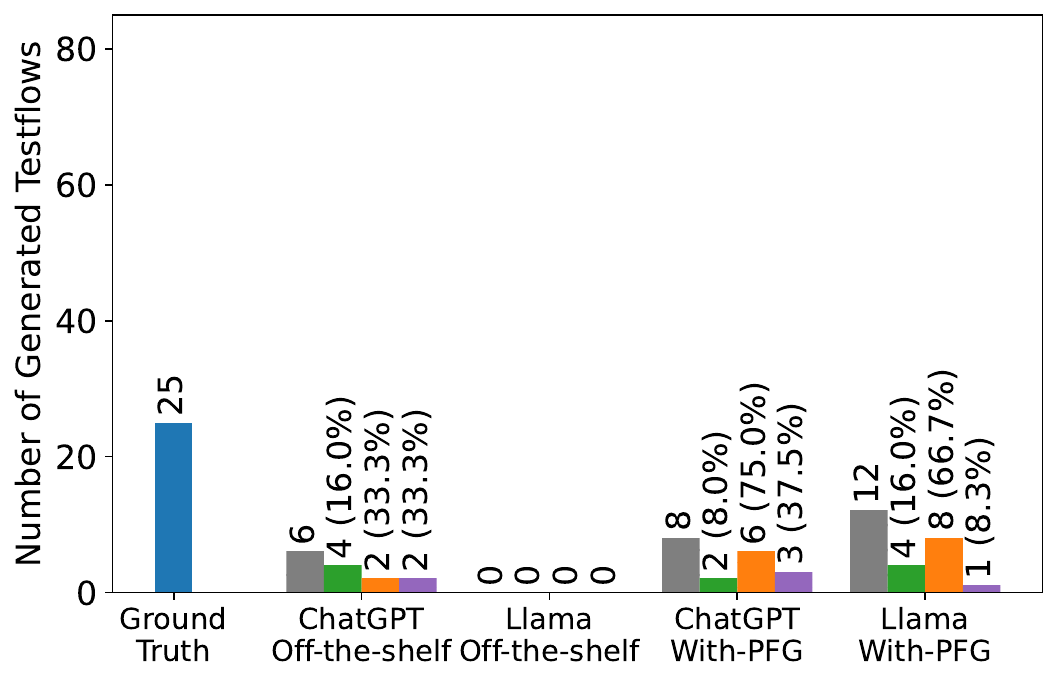}    
    }
    \caption{ \ahmadou Comparison of testflows generated by Off-the-Shelf models and \sol's across five APT41 reports.}
    \label{fig:off-the-shelf_models_using_PFG}
    \vspace{-4mm}
\end{figure*}

Fig.~\ref{fig:fg_against_off_the_shelf} shows the comparison results of the testflows generated by \sol and Off-the-Shelf models across the five APT41 reports. We can see that \sol generates more than twice the number of valid testflows compared to {\tt ChatGPT-off-the-shelf} and {\tt Llama-off-the-shelf} across all reports. This demonstrates \sol's superior ability to produce valid testflows and underscores the significance of fine-tuning \sol with a customized NER model. 
In contrast, general-purpose LLMs, such as {\tt ChatGPT-off-the-shelf} and {\tt Llama-off-the-shelf}, lack accuracy in this specialized context.
More specifically, \sol consistently achieves higher coverage rates compared to off-the-shelf models. For example, in Mandiant \#1 and Mandiant \#2, \sol achieves nearly three times the coverage of {\tt ChatGPT-off-the-shelf} and {\tt Llama-off-the-shelf}. However, in Mandiant \#3, Mandiant \#4, and TrendMicro \#5, the coverage rates are almost identical across all models. These results highlight that \sol not only generates the highest number of testflows but also aligns closely with the practices of manual practitioners. 
%
%
While \sol covers the majority of testflows, it also demonstrates the ability to discover new valid testflows. \sol identifies new testflows nearly twice as often as the off-the-shelf models for most reports, except for Mandiant \#2, where {\tt Llama-off-the-shelf} discovers more new testflows than both \sol and {\tt ChatGPT-off-the-shelf}. This exception is justified by the fact that, for this particular report, \sol generated a higher number of covered tests, resulting in fewer opportunities for discovering new tests.}

\noindent Although \sol excels in generating valid, covered, and newly discovered testflows, its primary strength lies in producing high-quality testflows.
\sol consistently generates almost twice as many high-quality testflows, except the TrendMicro \#5 report. 
{\ahmadou Additionally, we observed that {\tt ChatGPT-off
-the-shelf} tends to generate testflows for which it has high confidence. Upon manually reviewing its output, many of the testflows were found to be nearly identical to the original text from the reports. While this approach ensures accuracy for individual testflows, it results in the omission of numerous testflows typically expected from an APT report. Consequently, the model produces very few testflows, rendering it less comprehensive for sophisticated attacks such as APT41\footnote{APT41: \url{https://attack.mitre.org/groups/G0096/}}.
On the other hand, {\tt Llama-off-the-shelf} generates more testflows compared to {\tt ChatGPT-off-the-shelf}. However, as shown in Figs.~\ref{fig:Mandiant_1-2019-08-07} and~\ref{fig:TrendMicro_2021-08-24}, there are instances where {\tt Llama-off-the-shelf} fails to produce any results. This limitation arises from the model's inability to effectively process certain reports, particularly those with extensive size or complexity. Even after splitting the input data into smaller sections, the model struggled to generate testflows for these challenging reports.}
{\ahmadou
\subsection{\sol's Pipeline with Off-the-Shelf Models}
This experiment demonstrates the added value of integrating our \sol pipeline into off-the-shelf models.
As shown in Fig.~\ref{fig:off-the-shelf_models_using_PFG}, the integration of our pipeline significantly enhances the capability of these models to generate a greater number of valid testflows. Specifically, the number of valid testflows produced by {\tt ChatGPT-off-the-shelf} and {\tt Llama-off-the-shelf} increased by at least fourfold in Mandiant \#1, \#2, and \#4. 
Moreover, even in Mandiant \#3 and TrendMicro \#5, where the off-the-shelf models struggled, the integration of \sol still led to a significant improvement.
This highlights the critical role of \sol in resolving logical inconsistencies and enriching the context with domain-specific entities. 

\noindent In report Mandiant \#1, the number of covered testflows generated by {\tt ChatGPT-off-the-shelf} increases from 5 to 17, and in report Mandiant \#2, from 6 to 27 -- more than quadrupling the original numbers. Similarly, for {\tt Llama-off-the-shelf}, coverage rises from 6 to 9. Reports Mandiant \#3, Mandiant \#4, and TrendMicro \#5 show moderate but consistent improvements, further demonstrating the significant impact of \sol on these models.
%
Similarly, the number of newly discovered testflows increased significantly, particularly for {\tt ChatGPT-off-the-shelf}, which saw a large rise, and for {\tt Llama-off-the-shelf}, which experienced a moderate increase. 
We also observed that the total number of high-quality testflows tripled for both {\tt ChatGPT-off-the-shelf} and {\tt Llama-off-the-shelf} in report Mandiant \#1, indicating improvement in the quality of testflows when \sol was integrated.
%

}

\begin{figure}[!h]
    \centering
    \includegraphics[width=.9\linewidth]{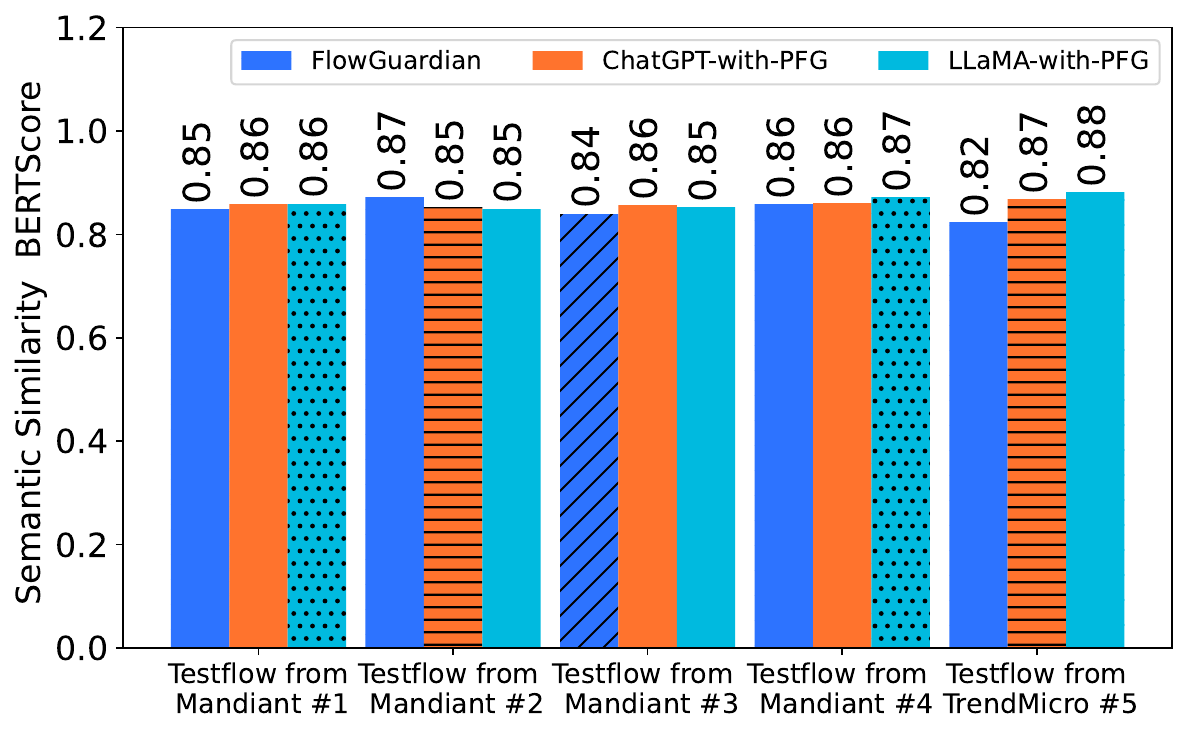}
    \caption{Semantic similarity of testflows generated by \sol and Off-the-Shelf models vs ground truth across five APT reports.}
    \label{fig:similarity_all_models}
\end{figure}

{\ahmadou
\subsection{Similarity Analysis between Generated Testflows}
\label{subsec:sim_testflow}
Fig.~\ref{fig:similarity_all_models} shows the semantic similarity scores between the testflows generated by \sol and the ground truth testflows across five reports. \sol demonstrates consistent semantic similarity across all reports, ranging from 0.82 to 0.87. This consistency highlights \sol's effectiveness in generating testflows that are semantically aligned with the ground truth, regardless of the report.
%
\texttt{ChatGPT-with-PFG} also shows consistent performance, with minor variations across the reports. 
However, in Mandiant \#2, a slight drop in the similarity score is observed. This decline may be attributed to the report's characteristics, such as ambiguity, less structure, or reliance on images. 
Conversely, the same figure
shows that \texttt{LLaMA-with-PFG} exhibits consistent performance across the reports, albeit with slightly higher variability compared to \texttt{ChatGPT-with-PFG}. Similar to \texttt{ChatGPT-with-PFG}, \texttt{LLaMA-with-PFG} appears to perform better with textual and structured reports.}


\subsection{Execution Time}
Manually extracting testflows involves thoroughly reading and analyzing the entire report, a process that typically takes several hours. 
{\revision For example, longer reports like Mandiant \#1 and TrendMicro \#5 are executed at 493.13 sec and 306.27 sec, respectively. 
In comparison to security practitioners, who take several hours/days, \sol's ability to handle such reports in under 10 minutes is a remarkable improvement. Shorter reports like Mandiant \#2 and Mandiant \#4 are executed at 35.9 sec and 82.29 sec, respectively. \sol processes them in less than two minutes, showing that \sol scales well across different report sizes and lengths. 
Medium reports that are neither too long nor too short show even good execution times. For example, in report Mandiant \#3, \sol can process the report in just 70.87 sec, which is far quicker than manual practitioners.
Overall, \sol demonstrates exceptional efficiency by reducing the analysis time from several hours to just minutes (average times 197.692 sec), depending on report size.
}
%
By drastically reducing processing time, \sol enables security teams to focus on higher-value analytical tasks rather than spending time on the labor-intensive creation of testflows, significantly enhancing operational efficiency.

\section{Conclusion \& Future Work}
\label{sec:conclusion}
This paper introduced \sol, an automated solution for attack testflows extraction from unstructured cyber threat reports. {\sepehr \sol employs a multi-component architecture that includes contextual analysis, sequence analysis, and testflow generation modules. 
The contextual analysis module leverages language models to understand the semantics and context of the extracted information, ensuring that relevant details are accurately captured. The sequence analysis component organizes these details into coherent sequences that reflect the logical flow of cyber attacks. Finally, the testflow generation module translates these sequences into actionable testflows that can be directly used by SOCs for threat hunting and incident response.}
Our evaluations indicate that \sol not only accelerates the extraction process but also enhances the quality of the resulting testflows. This advancement enables the SOC team to improve its threat hunting and incident response capabilities by providing structured, actionable intelligence. 
For future work, 
we intend to integrate \sol with existing SOC tools and workflows to provide dynamic testflows at run-time, facilitating seamless adoption by practitioners and enhancing operational efficiency. 


\bibliographystyle{IEEEtran}
\bibliography{Ref}

\begin{thebibliography}{10}
\providecommand{\url}[1]{#1}
\csname url@samestyle\endcsname
\providecommand{\newblock}{\relax}
\providecommand{\bibinfo}[2]{#2}
\providecommand{\BIBentrySTDinterwordspacing}{\spaceskip=0pt\relax}
\providecommand{\BIBentryALTinterwordstretchfactor}{4}
\providecommand{\BIBentryALTinterwordspacing}{\spaceskip=\fontdimen2\font plus
\BIBentryALTinterwordstretchfactor\fontdimen3\font minus \fontdimen4\font\relax}
\providecommand{\BIBforeignlanguage}[2]{{%
\expandafter\ifx\csname l@#1\endcsname\relax
\typeout{** WARNING: IEEEtran.bst: No hyphenation pattern has been}%
\typeout{** loaded for the language `#1'. Using the pattern for}%
\typeout{** the default language instead.}%
\else
\language=\csname l@#1\endcsname
\fi
#2}}
\providecommand{\BIBdecl}{\relax}
\BIBdecl

\bibitem{alshamrani2019survey}
A.~Alshamrani \emph{et~al.}, ``{A Survey on Advanced Persistent Threats: Techniques, Solutions, Challenges, and Research Opportunities},'' \emph{IEEE COMST}, vol.~21, no.~2, pp. 1851--1877, 2019.

\bibitem{googleblog2024}
{Google Cloud}, ``{APT41 Initiates Global Intrusion Campaign Using Multiple Exploits},'' 2024.

\bibitem{googleblog2023}
{Mandiant - Google Cloud}, ``{Game Over: Detecting and Stopping an APT41 Operation},'' 2023.

\bibitem{mandiant2024}
{Mandiant - Google Cloud}, ``{Double Dragon APT41, a dual espionage and cyber crime operation},'' 2022.

\bibitem{gcat2023apr}
{Google Cybersecurity Action Team}, ``{Threat Horizons - April 2023 Threat Horizons Report},'' 2023.

\bibitem{li2024automated}
L.~Li \emph{et~al.}, ``{Automated discovery and mapping ATT\&CK tactics and techniques for unstructured cyber threat intelligence},'' \emph{Computers \& Security}, vol. 140, 2024.

\bibitem{abdeen2023smet}
B.~Abdeen \emph{et~al.}, ``Smet: Semantic mapping of cve to att\&ck and its application to cybersecurity,'' in \emph{IFIP DBSec}, 2023.

\bibitem{wiliam2024interview}
W.~P.~M. III \emph{et~al.}, ``{An Interview Study on Third-Party Cyber Threat Hunting Processes in the U.S. Department of Homeland Security},'' in \emph{USENIX Security}, 2024.

\bibitem{10.1145/3607199.3607208}
M.~T. Alam \emph{et~al.}, ``{Looking Beyond IoCs: Automatically Extracting Attack Patterns from External CTI},'' in \emph{RAID}, 2023.

\bibitem{devlin2018bert}
J.~Devlin \emph{et~al.}, ``{BERT: Pre-training of Deep Bidirectional Transformers for Language Understanding},'' \emph{arXiv}, 2018.

\bibitem{liu2019roberta}
Y.~Liu \emph{et~al.}, ``{Roberta: A robustly optimized BERT pretraining approach},'' \emph{arXiv}, 2019.

\bibitem{nayak2021mum}
P.~Nayak, ``{MUM: A new AI milestone for understanding information},'' \emph{Google}, vol.~18, 2021.

\bibitem{motlagh2024largelanguagemodelscybersecurity}
F.~N. Motlagh \emph{et~al.}, ``Large language models in cybersecurity: State-of-the-art,'' \emph{arXiv}, 2024.

\bibitem{bhusal2024securebenchmarkinggenerativelarge}
D.~Bhusal \emph{et~al.}, ``Secure: Benchmarking generative large language models for cybersecurity advisory,'' \emph{arXiv}, 2024.

\bibitem{guo2023framework}
Y.~Guo \emph{et~al.}, ``A framework for threat intelligence extraction and fusion,'' \emph{Computers \& Security}, vol. 132, p. 103371, 2023.

\bibitem{9946567}
N.~Prayogo \emph{et~al.}, ``Context-aware attended-over distributed specificity for information extraction in cybersecurity,'' in \emph{IEEE IEMCON}, 2022.

\bibitem{10266600}
H.~Belani \emph{et~al.}, ``Ontology-based cybersecurity for well-being, aging and health: A scoping review,'' in \emph{IEEE MeditCom}, 2023.

\bibitem{9526808}
P.~Abdurehim \emph{et~al.}, ``{A short review of relation extraction methods},'' in \emph{ICICTA}, 2020.

\bibitem{satvat2021extractor}
K.~Satvat \emph{et~al.}, ``{Extractor: Extracting attack behavior from threat reports},'' in \emph{IEEE EuroS\&P}.\hskip 1em plus 0.5em minus 0.4em\relax IEEE, 2021.

\bibitem{kremer2023ic}
R.~Kremer \emph{et~al.}, ``{IC-SECURE: Intelligent System for Assisting Security Experts in Generating Playbooks for Automated Incident Response},'' \emph{arXiv}, 2023.

\bibitem{gao2021enabling}
P.~Gao \emph{et~al.}, ``Enabling efficient cyber threat hunting with cyber threat intelligence,'' in \emph{IEEE ICDE}.\hskip 1em plus 0.5em minus 0.4em\relax IEEE, 2021.

\bibitem{husari2017ttpdrill}
G.~Husari \emph{et~al.}, ``{TTPDrill: Automatic and Accurate Extraction of Threat Actions from Unstructured Text of CTI Sources},'' in \emph{ACSAC}, 2017.

\bibitem{10128641}
B.~H.~S. Durga \emph{et~al.}, ``Information extraction from text messages using natural language processing,'' in \emph{ICCCI}, 2023, pp. 1--6.

\bibitem{10428218}
Z.~Qiao \emph{et~al.}, ``Improving cybersecurity named entity recognition with large language models,'' in \emph{CSECS}, 2023.

\bibitem{googleblogAPT41states}
{Google Cloud}, ``Apt41 us state governments: Threat intelligence insights,'' 2023.

\bibitem{saadany2021bleu}
H.~Saadany \emph{et~al.}, ``Bleu, meteor, bertscore: evaluation of metrics performance in assessing critical translation errors in sentiment-oriented text,'' \emph{arXiv}, 2021.

\bibitem{trendmicro2021earth}
{Trend Micro}, ``{Earth Baku An APT Group Targeting Indo-Pacific Countries With New Stealth Loaders and Backdoor},'' 2021.

\end{thebibliography}

\end{document}